\documentclass[reprint,superscriptaddress,amsmath,amssymb,aip,jcp,floatfix]{revtex4-2}

\usepackage{graphicx}
\usepackage{dcolumn}
\usepackage{bm}
\usepackage{booktabs}
\usepackage{multirow}
\usepackage[hypertexnames=false,hidelinks]{hyperref}
\hypersetup{
  pdftitle={From Energy--Force Weighting to Primal--Dual Optimization of Machine-Learned Interatomic Potentials},
  pdfauthor={Chenyu Wang, Yangshuai Wang, and Lei Zhang}
}
\usepackage[capitalise,noabbrev]{cleveref}
\usepackage{algorithm}
\usepackage[noend]{algpseudocode}
\usepackage{siunitx}
\usepackage{mathrsfs}
\usepackage{amsmath, amssymb, amsthm}

\numberwithin{equation}{section}
\begin{document}

\title{From Energy--Force Weighting to Primal--Dual Optimization of Machine-Learned Interatomic Potentials}

\author{Chenyu Wang}
\affiliation{School of Mathematical Sciences, Institute of Natural Sciences and MOE-LSC, Shanghai Jiao Tong University, Shanghai 200240, China.}

\author{Yangshuai Wang}
\email{yswang@nus.edu.sg}
\affiliation{Department of Mathematics, National University of Singapore, 10 Lower Kent Ridge Road, Singapore.}

\author{Lei Zhang}
\affiliation{School of Mathematical Sciences, Institute of Natural Sciences and MOE-LSC, Shanghai Jiao Tong University, Shanghai 200240, China.}

\date{1 October 2026}

\begin{abstract}
Machine-learned interatomic potentials are commonly fitted by weighted-sum scalarization, which combines energy and force errors in a single loss. A nominal weight, however, identifies a potential only relative to the complete fitting protocol. We therefore treat energy--force balancing as a protocol-dependent problem of physical model selection. For fixed-basis atomic cluster expansion models of molten LiCl, liquid H$_2$O, and Si, the resulting scalarization paths contain dominated states at extreme weights. Their nondominated subsets depend on the solver, and the out-of-distribution Si path is nonmonotone. We replace direct weight selection by minimizing the regularized energy objective subject to an upper bound on the normalized force loss. Projected dual ascent adjusts the Lagrange multiplier from the force-constraint residual. A frozen-multiplier limited-memory quasi-Newton refinement then returns the final model. Across the three systems, the constrained procedure reaches the solver-matched low-error region and gives measured speedups of $6.7$--$8.9$ over completed scalarization scans under the stated timing convention. Physical-property calculations show that first-shell geometry is comparatively insensitive to the selected balance, whereas transport and solid-state observables vary more strongly. These results identify the fitted model--protocol pair as the relevant object of energy--force model selection and support force-constrained fitting as an explicit selection rule.
\end{abstract}

\maketitle

\section{Introduction}
\label{sec:intro}

Machine-learned interatomic potentials (MLIPs) approximate the Born--Oppenheimer potential-energy surface for molecular dynamics, structure relaxation, and response calculations \cite{deringer2019machine,behler2021four}. A representation and a reference dataset define a family of candidate surfaces, but the training protocol determines which member is ultimately used in simulation. Most supervised MLIP fits combine energy and force losses in a weighted sum. This is a form of scalarization, which reformulates a multiobjective optimization problem as a single-objective problem. This construction spans descriptor-based potentials \cite{behler2007generalized,bartok2010gaussian,smith2017ani,chmiela2017machine}, systematically improvable basis expansions \cite{drautz2019atomic,shapeev2016moment,thompson2015spectral}, and message-passing or E(3)-equivariant architectures \cite{schutt2017schnet,gasteiger2020directional,schutt2021equivariant,batzner2022,musaelian2023learning,batatia2022mace}. Across these models, the energy--force balance determines which fitted surface the protocol returns and thereby becomes part of physical model selection.

Energy and force labels constrain complementary aspects of the same differentiable surface. Energies determine relative stability and thermodynamic ordering, whereas forces encode the local derivatives that govern motion and response. An imbalanced fit can distort liquid transport or solid-state response even when one fitting error remains small. Greater force emphasis may improve local derivatives at the expense of energetic ordering, while an energy-biased fit can leave large force errors. This compromise governs which features of the potential-energy surface the fitted model preserves most accurately.

The numerical meaning of a nominal weight and the accessible energy--force tradeoffs depend on the complete training protocol, including loss normalization, regularization, data partitioning, initialization, optimization, and stopping. The same weight can select different potential-energy surfaces under different protocols and is meaningful only relative to the protocol that maps it to a fitted model.

Existing MLIP training strategies either prescribe the energy--force coefficient or adapt it during optimization. Prescribed choices include fixed scalings in NequIP-style protocols \cite{batzner2022}, optimized fractions in atomic cluster expansions \cite{bochkarev2022efficient,lysogorskiy2021performant}, and property-specific weights in SNAP fits \cite{wood2018extending}; direct training guidelines discuss their practical selection \cite{tokita2023how}. DeePMD-style schedules \cite{zhang2018deep,WANG2018} and adaptive MLIP weighting \cite{OCAMPO2024} vary the coefficient during optimization, while multi-task and multiobjective methods introduce more general balancing mechanisms \cite{kendall2018multi,chen2018gradnorm,desideri2012multiple,sener2018multi}. These approaches determine how the coefficient is assigned or updated, but its meaning remains tied to the full training protocol. The unresolved problem is the selection of a fitted potential under a specified protocol and a stated force-loss tolerance.

Pareto analysis provides a natural way to expose this protocol dependence. Related analyses have examined competing objectives in physics-informed learning \cite{rohrhofer2023data} and application-specific MLIP design \cite{baghishov2026application}. For a fixed training protocol, we map each sampled scalarization weight to the held-out energy and force losses of the returned model. In the systems studied here, the resulting paths include dominated extreme-weight states and nonmonotone segments; their empirical nondominated subsets also depend on the solver. We use each sampled path as a protocol-specific comparison set. Constructing this set requires a separate fit at every weight, followed by an additional criterion for selecting a compromise.

\begin{figure*}[t]
    \centering
    \includegraphics[width=\linewidth]{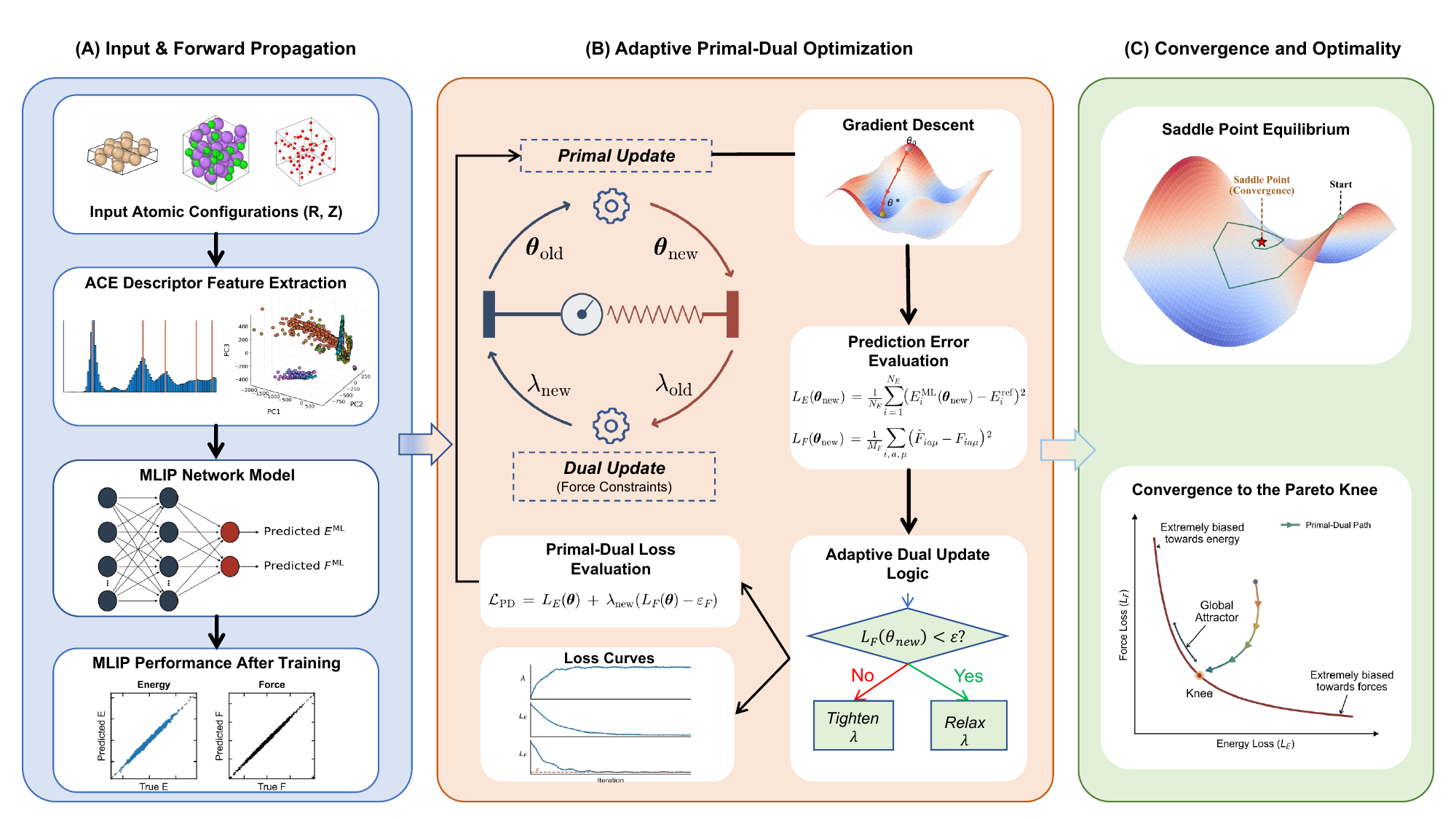}
    \caption{Protocol-dependent energy--force model selection. (a) A differentiable MLIP maps reference configurations to energy and force predictions, which determine fitting errors and downstream physical observables. The present calculations use a fixed-basis atomic cluster expansion. (b) Primal updates reduce the current Lagrangian, while projected dual updates adjust the multiplier from the force-constraint residual. (c) The resulting trajectory is compared with the low-error region of the empirical Pareto subset before the selected model is evaluated through physical observables. Secs.~\ref{sec:formulation}--\ref{sec:ace_linear_relation} define the objectives, multiplier update, and discrete reference set.}
    \label{fig:schematic_overview}
\end{figure*}

We replace direct weight tuning with force-constrained model selection. For fixed-basis atomic cluster expansion (ACE) models of molten LiCl, liquid H$_2$O, and Si, we minimize the regularized energy objective subject to an upper bound on the normalized squared force objective. The adaptive stage alternates primal descent with projected dual ascent, which updates the associated multiplier from the constraint residual. The energy--force coefficient is thus treated as the dual variable associated with a stated force-loss tolerance rather than as a prescribed hyperparameter. Freezing the returned multiplier defines the objective for deterministic limited-memory Broyden--Fletcher--Goldfarb--Shanno (L-BFGS) refinement, which returns the final model. The fixed-basis ACE setting isolates balance selection from representation learning. At the objective level, the same construction can be formulated whenever the required energy- and force-loss gradients are available. A related exact-gradient result motivates time-scale separation in the implemented Adam--L-BFGS iteration. Figure~\ref{fig:schematic_overview} summarizes model fitting, multiplier adaptation, and physical evaluation.

Together, molten LiCl, liquid H$_2$O, and Si span ionic, molecular, and covalent environments, allowing us to trace energy--force selection from held-out errors to physical observables. Across all three systems, the constrained fits remain within $2.42\%$ of their solver-matched reference RMSEs and yield measured speedups of $6.7$--$8.9$ over completed sampled grids under the stated timing convention. Distinct LiCl warm starts reach the same low-error region with separate endpoints, showing that multiplier feedback reduces initialization sensitivity in these fits. The downstream calculations reveal observable-dependent sensitivity: first-shell geometry changes little, while transport and solid-state observables show greater model-to-model variation.

\section{Methods}

We distinguish the physical RMSEs used for evaluation from the scaled squared objectives used for fitting. This section first defines the empirical scalarization path returned by a complete training protocol, then replaces direct weight selection with a force-loss constraint and derives the corresponding multiplier update for fixed-basis ACE models. The formulation retains protocol dependence while supplying an explicit selection rule within a fully specified fitting procedure.

\subsection{Problem formulation}
\label{sec:formulation}

Let $\mathbf{R}$ and $\mathbf{Z}$ denote the atomic coordinates and corresponding atomic numbers, respectively. An MLIP represents the Born--Oppenheimer potential-energy surface by a differentiable total-energy function $E^{\mathrm{ML}}=\mathcal{E}(\mathbf{R},\mathbf{Z};\boldsymbol{\theta})$ with trainable parameters $\boldsymbol{\theta}$. The predicted forces are its negative coordinate gradient,
$\mathbf{F}^{\mathrm{ML}}=-\nabla_{\mathbf{R}}\mathcal{E}(\mathbf{R},\mathbf{Z};\boldsymbol{\theta})$, which couples energy and force predictions by construction. Differentiating a force loss with respect to $\boldsymbol{\theta}$ therefore requires mixed coordinate--parameter derivatives of $\mathcal{E}$. These derivatives are analytic for the fixed-basis ACE models used here and are typically obtained by automatic differentiation for nonlinear MLIPs.

The training data comprise $N$ reference configurations,
$\mathcal{D}=\{(\mathbf{R}_i,\mathbf{Z}_i,E_i^{\mathrm{ref}},\mathbf{F}_i^{\mathrm{ref}})\}_{i=1}^{N}$,
with reference total energies and atomic forces obtained from electronic-structure calculations. Let $\mathcal{I}_E$ and $\mathcal{I}_F$ index the configurations with energy and force labels, respectively. Let $N_E=|\mathcal{I}_E|$, $N_i$ be the number of atoms in configuration $i$, and $M_F=\sum_{i\in\mathcal{I}_F}3N_i$ be the total number of labeled scalar force components. To distinguish physical-scale errors from the scaled fitting objectives introduced below, we define the unweighted mean-squared errors (MSEs) underlying joint energy--force fitting in force matching \cite{ercolessi1994} and modern MLIP training \cite{unke2021}
\begin{align}
J_E(\boldsymbol{\theta}) &= \frac{1}{N_E}\sum_{i\in\mathcal{I}_E}\big(E_i^{\mathrm{ML}} - E_i^{\mathrm{ref}}\big)^2, \nonumber \\
J_F(\boldsymbol{\theta}) &= \frac{1}{M_F}\sum_{i\in\mathcal{I}_F}\sum_{a=1}^{N_i}\sum_{\mu=1}^{3}
\big(F_{ia\mu}^{\mathrm{ML}} - F_{ia\mu}^{\mathrm{ref}}\big)^2.
\label{eq:energy_force_loss}
\end{align}
The index sets are kept separate to allow a fitting protocol to filter energy and force labels differently, although every configuration used here carries both labels. The optimizer acts on numerical objectives $L_E$ and $L_F$ that incorporate unit conversion, per-configuration or per-atom weighting, label-count normalization, and package-level scaling, instead of directly using $J_E$ and $J_F$. Section~\ref{sec:ace_linear_relation} gives their ACE realization. Once these choices are fixed, $L_E$ and $L_F$ define the two data-fit coordinates of the empirical scalarization path. A fixed differentiable regularizer $\mathcal{R}(\boldsymbol{\theta})$ affects the returned model and remains outside these two Pareto coordinates. A regularized weighted-sum scalarization with nominal energy fraction $\alpha\in[0,1]$ is
\begin{equation}
\mathcal{L}(\boldsymbol{\theta};\alpha) = \alpha L_E(\boldsymbol{\theta}) + (1-\alpha)L_F(\boldsymbol{\theta}) + \mathcal{R}(\boldsymbol{\theta}).
\label{eq:alpha_loss}
\end{equation}
Equation~\eqref{eq:alpha_loss} defines the scalarized objective, while the initialization, optimizer, and stopping rule complete the map to a returned fitted state.

Throughout, $L_E$ and $L_F$ denote the scaled squared objectives specified by the training protocol, whereas numerical tables report errors in physical units. For an evaluation configuration $\mathbf{R}$ containing $N_{\mathbf{R}}$ atoms, define $\Delta E(\mathbf{R})=E^{\mathrm{ML}}(\mathbf{R})-E^{\mathrm{ref}}(\mathbf{R})$ and $\Delta F_{a\mu}(\mathbf{R})=F_{a\mu}^{\mathrm{ML}}(\mathbf{R})-F_{a\mu}^{\mathrm{ref}}(\mathbf{R})$. We report a configuration-averaged per-atom energy RMSE and a component-averaged force RMSE as
\begin{align}
\mathrm{RMSE}_E
&=
10^3\left\langle
\left(\frac{\Delta E(\mathbf{R})}{N_{\mathbf{R}}}\right)^2
\right\rangle_{\mathbf{R}}^{1/2}
\ \mathrm{meV/atom},
\nonumber \\
\mathrm{RMSE}_F
&=
10^3
\left\langle
|\Delta F_{a\mu}(\mathbf{R})|^2
\right\rangle_{\mathbf{R},a,\mu}^{1/2}
\ \mathrm{meV}/\text{\AA}.
\label{eq:reported_rmse}
\end{align}
Here $\langle\cdot\rangle_{\mathbf{R}}$ denotes an equal average over configurations and $\langle\cdot\rangle_{\mathbf{R},a,\mu}$ an average over scalar force components. When energies and forces are expressed in eV and eV/\AA, respectively, the prefactor $10^3$ converts the reported errors to meV/atom and meV/\AA.

For a specified data split, $L_{E,\mathrm{train}}$ and $L_{F,\mathrm{train}}$ denote the fitting objectives, while $L_{E,\mathrm{test}}$ and $L_{F,\mathrm{test}}$ denote their counterparts on the held-out split. We retain the dataset label ``test'' for the split that constructs the empirical scalarization path and selects its discrete knee. It serves as the model-selection split throughout the paper. Accordingly, $\alpha$ sets the energy--force balance only within the fully specified model-selection protocol.

\subsection{Empirical scalarization path and Pareto subset}
\label{sec:pareto}

Equation~\eqref{eq:alpha_loss} scalarizes the data-fit vector
\begin{equation}
    \mathbf{L}(\boldsymbol{\theta}) = \begin{bmatrix} L_E(\boldsymbol{\theta}) \\ L_F(\boldsymbol{\theta}) \end{bmatrix}.
\end{equation}
Let $\mathbf{L}_{\mathrm{test}}$ denote its held-out counterpart. We use this vector to compare models returned by a fixed training protocol; unless stated otherwise, dominance comparisons are based on its two coordinates rather than on the scalarized training objective.

We identify nondominated points using the standard multiobjective dominance relation \cite{sener2018multi}. On an evaluation split $s\in\{\mathrm{train},\mathrm{test}\}$, a parameter vector $\boldsymbol{\theta}_1$ dominates $\boldsymbol{\theta}_2$, written $\boldsymbol{\theta}_1\prec_s\boldsymbol{\theta}_2$, if
\begin{equation}
    \begin{aligned}
        L_{k,s}(\boldsymbol{\theta}_1) &\le L_{k,s}(\boldsymbol{\theta}_2) \quad \text{for all } k \in \{E, F\}, \\
        L_{k,s}(\boldsymbol{\theta}_1) &< L_{k,s}(\boldsymbol{\theta}_2) \quad \text{for at least one } k \in \{E, F\}.
    \end{aligned}
\end{equation}
The empirical construction below applies this relation with $s=\mathrm{test}$.

Let $\Theta$ be the model parameter space. The ideal model-space Pareto set on split $s$ is
\begin{equation}
    \mathcal{P}^*_s =
    \{ \boldsymbol{\theta}\in\Theta \mid \nexists \, \boldsymbol{\theta}'\in\Theta \text{ such that }
    \boldsymbol{\theta}' \prec_s \boldsymbol{\theta} \}.
\end{equation}
Its image in objective space is the ideal Pareto front $\mathcal{PF}^*_s$. Numerical training provides a more restricted object: the models returned at sampled weights. Following the data-driven apparent-Pareto formulation of Rohrhofer \textit{et al.} \cite{rohrhofer2023data}, we construct the empirical set from these returned models. Let $\mathcal{T}$ specify the basis and data split, the loss scaling and regularization, and the initialization, stochastic realization, optimizer, and stopping criterion. For each sampled $\alpha$, define the returned state by
\begin{equation}
    \boldsymbol{\theta}^{\mathcal{T}}(\alpha)
    =
    \mathrm{Train}_{\mathcal{T}}(\alpha).
\end{equation}
Regardless of whether the minimizer is unique or the optimizer reaches global optimality, the empirical path is defined by $\boldsymbol{\theta}^{\mathcal{T}}(\alpha)$ rather than by an abstract global minimizer. The same returned model is used for held-out and physical evaluation.

Let $\mathcal{A}$ denote the finite scalarization grid. The empirical scalarization path in held-out objective space is
\begin{equation}
\mathcal{S}_{\mathrm{app}}
=
\big\{\mathbf{L}_{\mathrm{test}}(\boldsymbol{\theta}^{\mathcal{T}}(\alpha))
\mid \alpha\in\mathcal{A}\big\}.
\label{eq:scalarization_path}
\end{equation}
Let $\mathcal{PF}_{\mathrm{app}}\subseteq\mathcal{S}_{\mathrm{app}}$ denote its nondominated subset. The corresponding sampled weights form
\begin{equation}
    \mathcal{A}_{\mathrm{nd}}
    =
    \{\alpha\in\mathcal{A}\mid
    \mathbf{L}_{\mathrm{test}}(\boldsymbol{\theta}^{\mathcal{T}}(\alpha))
    \in \mathcal{PF}_{\mathrm{app}}\}.
\end{equation}
The full path can contain dominated sampled points.

The subscript ``app'' records finite sampling and dependence on the training protocol. Even though each fixed-weight ACE subproblem is convex, projecting the returned models into held-out coordinates can produce an irregular or nonmonotone path. Removing dominated samples leaves the protocol dependence unchanged, and a finite linear-scalarization grid can miss unsampled supported states and unsupported nondominated states. We use $\mathcal{PF}_{\mathrm{app}}$ as a finite, protocol-specific comparison set.

From each empirical Pareto subset, we select a reference knee by applying a maximum chord-distance criterion inspired by Kneedle\cite{satopaa2011finding} in normalized logarithmic held-out loss coordinates. The logarithm compresses the several orders of magnitude spanned by extreme scalarizations, while coordinatewise min--max normalization places the energy and force losses on a common scale. Write $L_{q,\mathrm{test}}(\alpha)=L_{q,\mathrm{test}}(\boldsymbol{\theta}^{\mathcal{T}}(\alpha))$. Because all losses on the reported paths are strictly positive, define the normalized logarithmic loss for $q \in \{E,F\}$ by
\begin{equation}
    \widetilde L_q^{\log}(\alpha)
    =
    \frac{
    \log_{10} L_{q,\mathrm{test}}(\alpha)
    -
    \log_{10} L_q^{\min}
    }{
    \log_{10} L_q^{\max}
    -
    \log_{10} L_q^{\min}
    },
\end{equation}
where, for each system, $L_q^{\min}$ and $L_q^{\max}$ are taken over $\alpha \in \mathcal{A}_{\mathrm{nd}}$. Both coordinate ranges are nonzero for the reported Pareto subsets. The energy- and force-minimizing endpoints map to $(0,1)$ and $(1,0)$, respectively, so the chord joining them satisfies $\widetilde L_E^{\log}+\widetilde L_F^{\log}=1$. Only points satisfying
\(\widetilde L_E^{\log}+\widetilde L_F^{\log}<1\)
lie on the utopia-facing side of the chord. We select the discrete knee as the sampled nondominated point farthest from this chord,
\begin{equation}
\begin{aligned}
    \alpha_{\mathrm{knee}}
    &= \arg\max_{\substack{
        \alpha\in\mathcal A_{\mathrm{nd}}\\
        \widetilde L_E^{\log}(\alpha)+\widetilde L_F^{\log}(\alpha)<1
    }}
    \\[-0.2em]
    &\quad
    \left[1-\left(
        \widetilde L_E^{\log}(\alpha)
        +\widetilde L_F^{\log}(\alpha)
    \right)\right].
\end{aligned}
    \label{eq:knee_selection}
\end{equation}
If no sampled nondominated point lies on the utopia-facing side of the chord, no chord-based knee is assigned. The maximizer is unique on each reported path. The selected knee is a coordinate-dependent reference point. Appendix~\ref{sec:appendix_quantitative} lists the selected states, and Appendix~\ref{sec:appendix_knee_sensitivity} records an alternative linear-scale RMSE construction.

Constructing $\mathcal{S}_{\mathrm{app}}$ requires one independent fit for each sampled $\alpha$. The resulting path supplies both the solver-matched reference knee and the low-error training-force-loss scale used to set a common numerical tolerance. After this tolerance is fixed, the constrained route follows a single parameter--multiplier trajectory rather than fitting every point on the scalarization grid.

\subsection{Constrained multiplier update via primal--dual formulation}
\label{sec:pd}

The reference knee is a post hoc choice among independently trained scalarization models. We replace this weight scan by minimizing the regularized energy objective subject to a prescribed upper bound on the force objective. In the deterministic constrained problem, the Lagrange multiplier acts as the force coefficient; the numerical iteration updates it from the force-constraint residual and projects it onto a bounded interval.

We impose the force-loss bound on the training objective,
\begin{equation}
    \min_{\boldsymbol{\theta}} \ L_{E,\mathrm{train}}(\boldsymbol{\theta})+\mathcal{R}(\boldsymbol{\theta})
    \quad \text{subject to} \quad 
    L_{F,\mathrm{train}}(\boldsymbol{\theta}) \le \varepsilon_F.
    \label{eq:primal_problem_pd}
\end{equation}
Problem~\eqref{eq:primal_problem_pd} minimizes the regularized training energy loss over the force-feasible set. A positive tolerance avoids requiring a finite model class to interpolate the reference forces exactly. Reference electronic-structure data can contain numerical and modeling errors associated with convergence thresholds, basis sets, grids, and exchange--correlation approximations \cite{lejaeghere2014error,lam2026cutting,kruse2012geometrical}.

Feasibility requires $\varepsilon_F \ge \inf_{\boldsymbol{\theta}}L_{F,\mathrm{train}}(\boldsymbol{\theta})$ for the chosen model class and protocol. If this necessary condition fails, Problem~\eqref{eq:primal_problem_pd} is infeasible. If an unconstrained minimizer of $L_{E,\mathrm{train}}+\mathcal{R}$ is feasible, an associated Karush--Kuhn--Tucker (KKT) multiplier can be zero. For the unweighted physical force MSE $J_F$ in Eq.~\eqref{eq:energy_force_loss}, a prescribed RMSE $\tau_F$ in meV/\AA{} corresponds to $(\tau_F/10^3)^2$ in $(\mathrm{eV}/\text{\AA})^2$. The ACE force objective $L_F$ additionally contains configuration weights and package scaling. Accordingly, $\varepsilon_F$ specifies a system- and protocol-specific tolerance in the scaled numerical objective.

Introducing a nonnegative multiplier $\lambda\ge 0$ gives the uncapped primal--dual (PD) Lagrangian \cite{bertsekas1982constrained}
\begin{equation}
    \mathcal{L}_{\mathrm{PD}}(\boldsymbol{\theta}, \lambda)
    = L_{E,\mathrm{train}}(\boldsymbol{\theta}) + \mathcal{R}(\boldsymbol{\theta}) + \lambda \big(L_{F,\mathrm{train}}(\boldsymbol{\theta}) - \varepsilon_F\big),
    \label{eq:lagrangian_pd}
\end{equation}
The associated saddle-point problem minimizes $\mathcal{L}_{\mathrm{PD}}$ with respect to $\boldsymbol{\theta}$ and maximizes it over $\lambda\ge 0$. Projected dual ascent updates $\lambda$ from the constraint residual $L_{F,\mathrm{train}}-\varepsilon_F$: the multiplier increases when the constraint is violated and decreases, subject to nonnegativity, when the force loss is below tolerance. In each dual increment, the tolerance contributes the additive term $-\eta_\lambda\varepsilon_F$. When $L_{F,\mathrm{train}}$ remains well above $\varepsilon_F$ during the early adaptive stage, the measured force loss dominates the update, which makes the trajectory weakly sensitive to moderate changes in the tolerance scale. The KKT conditions below refer to the uncapped problem, whereas the implemented iteration projects the multiplier onto $[0,\Lambda_{\max}]$. An iterate with $\lambda=\Lambda_{\max}$ is therefore governed by the numerical upper bound.

For a fixed multiplier, the constant term $-\lambda\varepsilon_F$ is independent of $\boldsymbol{\theta}$, so the primal subproblem is equivalent to minimizing
\begin{equation}
    L_{E,\mathrm{train}}(\boldsymbol{\theta}) + \lambda L_{F,\mathrm{train}}(\boldsymbol{\theta}) + \mathcal{R}(\boldsymbol{\theta}).
    \label{eq:frozen_lambda_objective}
\end{equation}
Normalizing these data-term coefficients by the positive factor $1+\lambda$ defines the effective energy and force fractions
\begin{equation}
    \alpha_{\mathrm{eff}}(\lambda) = \frac{1}{1+\lambda},
    \qquad
    1-\alpha_{\mathrm{eff}}(\lambda) = \frac{\lambda}{1+\lambda}.
    \label{eq:lambda_alpha_map}
\end{equation}
The same normalization gives
\begin{equation}
    \alpha_{\mathrm{eff}}L_{E,\mathrm{train}}
    +(1-\alpha_{\mathrm{eff}})L_{F,\mathrm{train}}
    +\frac{\mathcal{R}}{1+\lambda}.
    \label{eq:lambda_normalized_objective}
\end{equation}
Equation~\eqref{eq:lambda_alpha_map} maps $\lambda$ exactly onto the relative data-term weights. For the full regularized objective in Eq.~\eqref{eq:alpha_loss}, the same normalization rescales the regularizer to $\mathcal{R}/(1+\lambda)$, so the static-$\alpha$ and frozen-$\lambda$ objectives differ in their effective regularization. Their final held-out losses are nevertheless evaluated in the same coordinates. Large $\lambda$ emphasizes the force term and drives $\alpha_{\mathrm{eff}}$ toward zero, whereas $\lambda=0$ recovers the energy-only data objective. In the second stage, $\lambda$ is held fixed and Eq.~\eqref{eq:frozen_lambda_objective} is minimized by L-BFGS.

The adaptive stage applies a primal descent step at every iteration and a projected dual-ascent step periodically, as summarized in Algorithm~\ref{alg:pd_opt}. For a scalar $x$, its projection onto the multiplier interval is $\Pi_{[0,\Lambda_{\max}]}(x)=\min\{\Lambda_{\max},\max(0,x)\}$.

\begin{algorithm}[H]
    \caption{Force-constrained primal--dual update for machine-learned interatomic potentials}
\label{alg:pd_opt}
\begin{algorithmic}[1]
\Require Training data $\mathcal{D}=\{(\mathbf{R}_i,\mathbf{Z}_i,E_i^{\mathrm{ref}},\mathbf{F}_i^{\mathrm{ref}})\}_{i=1}^N$.
\Require Force-loss tolerance $\varepsilon_F$, multiplier cap $\Lambda_{\max}$, and initial state $(\boldsymbol{\theta}_0,\lambda_0)$ with $\lambda_0\in[0,\Lambda_{\max}]$.
\Require Primal optimizer $\mathrm{Optim}_{\eta_\theta}$, dual step size $\eta_\lambda$, and dual update period $K_{\text{dual}}$.
\Require Maximum iterations $T$ and a prescribed stopping predicate $\mathrm{Stop}_F$ applied at the periodic full-training force-objective evaluations used for dual updates.

\vspace{0.1cm}
\State Set $t_{\mathrm{last}}\leftarrow 0$
\For{iteration $t=0, 1, \dots, T-1$}
  \State Sample a mini-batch $\mathcal{B} \subseteq \mathcal{D}$

  \Statex \textit{Forward pass and primal gradient}
  \State Compute mini-batch predictions $\hat{E}(\boldsymbol{\theta}_t)$ and $\hat{\mathbf{F}}(\boldsymbol{\theta}_t) = -\nabla_{\mathbf{R}} \hat{E}(\boldsymbol{\theta}_t)$
  \State Evaluate $L_E^{\mathcal{B}}$ and $L_F^{\mathcal{B}}$ on $\mathcal{B}$ using the same scaling as the full objectives
  \State Compute parameter gradient $\mathbf{g}_t \leftarrow \nabla_{\boldsymbol{\theta}} L_E^{\mathcal{B}}(\boldsymbol{\theta}_t) + \nabla_{\boldsymbol{\theta}}\mathcal{R}(\boldsymbol{\theta}_t) + \lambda_t \nabla_{\boldsymbol{\theta}} L_F^{\mathcal{B}}(\boldsymbol{\theta}_t)$

  \Statex \textit{Primal descent update}
  \State Apply primal optimizer step: $\boldsymbol{\theta}_{t+1} \leftarrow \mathrm{Optim}_{\eta_\theta}(\boldsymbol{\theta}_t, \mathbf{g}_t)$

  \Statex \textit{Dual ascent update}
  \If{$\text{mod}(t+1,\, K_{\text{dual}}) = 0$}
    \State Evaluate the post-step residual on the full training set: $v_t \leftarrow L_{F,\mathrm{train}}(\boldsymbol{\theta}_{t+1}) - \varepsilon_F$
    \State Projected dual ascent: $\lambda_{t+1} \leftarrow \Pi_{[0,\Lambda_{\max}]}(\lambda_t+\eta_\lambda v_t)$
  \Else
    \State Hold multiplier constant: $\lambda_{t+1} \leftarrow \lambda_t$
  \EndIf
  \State $t_{\mathrm{last}}\leftarrow t+1$

  \Statex \textit{Full-training stopping test}
  \If{$\text{mod}(t+1,\, K_{\text{dual}}) = 0$ and $\mathrm{Stop}_F$ is satisfied by the full-training evaluation above}
    \State \textbf{break}
  \EndIf
\EndFor
\vspace{0.1cm}
\Ensure Adaptive-stage state $(\boldsymbol{\theta}_{\mathrm{out}},\lambda_{\mathrm{out}}) \leftarrow (\boldsymbol{\theta}_{t_{\mathrm{last}}},\lambda_{t_{\mathrm{last}}})$ for subsequent fixed-$\lambda$ refinement
\end{algorithmic}
\end{algorithm}

The mini-batch losses in Algorithm~\ref{alg:pd_opt} use the same scaling as their full-training counterparts for the primal update. Every $K_{\mathrm{dual}}$ primal steps, the dual update instead evaluates the full-training constraint residual $L_{F,\mathrm{train}}(\boldsymbol{\theta}_{t+1})-\varepsilon_F$. The same scheduled full-training evaluation supplies the stopping quantity $\mathrm{Stop}_F$. After the adaptive stage, $\lambda_{\mathrm{out}}$ is frozen and $\boldsymbol{\theta}_{\mathrm{out}}$ initializes L-BFGS minimization of Eq.~\eqref{eq:frozen_lambda_objective} on the full training split.

For the fixed-basis ACE models considered here, both $\hat{E}$ and $\hat{\mathbf{F}}$ are linear in the fitted coefficients, so $\nabla_{\boldsymbol{\theta}}L_F$ is obtained analytically from the corresponding design matrices. For a nonlinear energy model, by contrast, the same gradient requires the mixed derivative $\nabla_{\boldsymbol{\theta}}\nabla_{\mathbf{R}}\hat{E}$, which can be evaluated by automatic differentiation. The KKT conditions below characterize the deterministic uncapped constrained problem. Final held-out losses evaluate the implemented two-stage procedure.

Suppose $\bar{\boldsymbol{\theta}}$ is a local solution of Eq.~\eqref{eq:primal_problem_pd}, the functions $L_{E,\mathrm{train}}$, $L_{F,\mathrm{train}}$, and $\mathcal{R}$ are differentiable, and a constraint qualification holds. Then there is an associated multiplier $\bar{\lambda}$ for which the KKT stationarity condition is
\begin{equation}
    \nabla_{\boldsymbol{\theta}} L_{E,\mathrm{train}}(\bar{\boldsymbol{\theta}})
    +
    \nabla_{\boldsymbol{\theta}}\mathcal{R}(\bar{\boldsymbol{\theta}})
    +
    \bar{\lambda}
    \nabla_{\boldsymbol{\theta}} L_{F,\mathrm{train}}(\bar{\boldsymbol{\theta}})
    =
    \mathbf{0},
\end{equation}
together with dual feasibility $\bar{\lambda}\ge 0$, primal feasibility $L_{F,\mathrm{train}}(\bar{\boldsymbol{\theta}})\le\varepsilon_F$, and complementarity $\bar{\lambda}(L_{F,\mathrm{train}}(\bar{\boldsymbol{\theta}})-\varepsilon_F)=0$ \cite{nocedal2006numerical}. These are necessary local optimality conditions under the stated assumptions. Because the fixed-basis ACE problem is convex, a KKT pair is also sufficient for global optimality, although the minimizing coefficient vector need not be unique. For the implemented procedure, $\lambda_{\mathrm{out}}$ is the force coefficient generated by the bounded update, and final held-out losses assess the subsequently refined model.

Appendix~\ref{sec:appendix_theory} relates the projected multiplier update to a simultaneous exact-gradient descent--ascent idealization with separated primal and dual step sizes \cite{lin2020gradient}. This exact-gradient result motivates the two-time-scale design. The implemented Adam--L-BFGS procedure is evaluated through final held-out losses relative to the solver-matched static references.

\subsection{Relation to fixed-basis ACE least squares and the multiplier iteration}
\label{sec:ace_linear_relation}

In the numerical experiments, the ACE basis functions are fixed and only the linear expansion coefficients are fitted, so $\boldsymbol{\theta}=\mathbf{c}$. This restriction isolates energy--force balance selection from concurrent representation learning. The stacked energy and force labels therefore take the schematic form
\begin{equation}
    \mathbf{y}_E \approx A_E \mathbf{c},
    \qquad
    \mathbf{y}_F \approx A_F \mathbf{c},
    \label{eq:ace_linear_system}
\end{equation}
where $A_E$ and $A_F$ are the energy and force design matrices. For fixed $\lambda$, fitting $\mathbf{c}$ reduces to the regularized weighted least-squares problem
\begin{align}
\min_{\mathbf{c}}\quad
&
\left\|W_E^{1/2}(A_E\mathbf{c}-\mathbf{y}_E)\right\|^2
+
\lambda
\left\|W_F^{1/2}(A_F\mathbf{c}-\mathbf{y}_F)\right\|^2
\nonumber\\
&+
\gamma\left\|\Gamma\mathbf{c}\right\|^2,
\label{eq:ace_weighted_ls}
\end{align}
where the positive semidefinite matrices $W_E$ and $W_F$ collect configuration weights, unit conversions, and normalization factors. The parameter $\gamma\ge0$ is the regularization strength, and $\Gamma$ denotes the ACE regularization operator. Equation~\eqref{eq:ace_weighted_ls} is therefore the fixed-basis ACE realization of the frozen-$\lambda$ objective $L_E^{\mathrm{ACE}}+\lambda L_F^{\mathrm{ACE}}+\mathcal{R}$, with $\mathcal{R}(\mathbf{c})=\gamma\|\Gamma\mathbf{c}\|^2$. For a configuration set $\mathcal{C}$, the component objectives in the \texttt{ACEpotentials.jl}-based fits take the form
\begin{align}
    L_E^{\mathrm{ACE}}(\mathbf{c}) &=
    \frac{1}{N_E}\sum_{\mathbf{R} \in \mathcal{C}} \omega_{\mathbf{R}}^E
    \big|E(\mathbf{c};\mathbf{R})-\mathscr{E}_{\mathbf{R}}\big|^2,
    \nonumber\\
    L_F^{\mathrm{ACE}}(\mathbf{c}) &=
    \frac{1}{M_F}\sum_{\mathbf{R} \in \mathcal{C}} \omega_{\mathbf{R}}^F
    \sum_{a=1}^{N_{\mathbf{R}}}\sum_{\mu=1}^{3}
    \big|F_{a\mu}(\mathbf{c};\mathbf{R})-\mathscr{F}_{\mathbf{R},a\mu}\big|^2,
    \nonumber\\
    M_F &= \sum_{\mathbf{R}\in\mathcal{C}}3N_{\mathbf{R}}.
    \label{eq:component-loss}
\end{align}
Every configuration in $\mathcal{C}$ carries both labels, so $N_E=|\mathcal{C}|$ and $M_F$ is the total number of labeled scalar force components. The quantities $\mathscr{E}_{\mathbf{R}}$ and $\mathscr{F}_{\mathbf{R},a\mu}$ are the reference \textit{ab initio} total energy and force components. The effective configuration weights $\omega_{\mathbf{R}}^E$ and $\omega_{\mathbf{R}}^F$ absorb the base configuration weighting, any per-atom energy normalization, unit conversions, and the fixed package-level scaling prescribed by the fitting protocol.

With $\gamma$ fixed, varying $\lambda$ changes the energy--force data-term ratio and, after normalization by $1+\lambda$, the effective regularization coefficient $\gamma/(1+\lambda)$, as shown in Eq.~\eqref{eq:lambda_normalized_objective}. For the fixed basis, Eq.~\eqref{eq:ace_weighted_ls} is a convex quadratic in $\mathbf{c}$. Joint rank deficiency of the weighted data and regularization operators permits multiple minimizing coefficient vectors; the numerical solver and stopping criterion then determine the returned vector. After $\lambda$ is selected, this convex subproblem can be refined deterministically on the full objective using a least-squares or quasi-Newton method.

For a fully assembled ACE problem, a deterministic alternative is to solve Eq.~\eqref{eq:ace_weighted_ls} repeatedly within a bracketed one-dimensional search over $\lambda$ for $L_{F,\mathrm{train}}(\mathbf{c}_\lambda)\approx\varepsilon_F$. If $\lambda_2>\lambda_1$ and $\mathbf{c}_{\lambda_1}$ and $\mathbf{c}_{\lambda_2}$ are exact global minimizers, summing the two optimality inequalities gives
$(\lambda_2-\lambda_1)[L_{F,\mathrm{train}}(\mathbf{c}_{\lambda_2})-L_{F,\mathrm{train}}(\mathbf{c}_{\lambda_1})]\le0$, so the optimal force objective is nonincreasing in $\lambda$. Each trial $\lambda$ requires a separately converged fit. The primal--dual iteration instead updates $\lambda$ along a single training trajectory before passing the returned value to fixed-$\lambda$ refinement. The reported comparisons assess the returned multiplier and refined model against the static ACE reference set. At the objective level, the constrained formulation and multiplier update extend to nonlinear MLIPs when the required mixed derivatives are available; the numerical experiments here use fixed-basis ACE.

\section{Results}
\label{sec:experiments}

The numerical study first examines how strongly the empirical energy--force trade-off depends on the training protocol and whether a force-constrained parameter--multiplier trajectory reaches the resulting solver-matched low-error region. We then assess how the selected balance affects liquid structure and transport and solid-state response. Across three chemical systems, we compare static L-BFGS paths, final constrained models, and downstream physical observables. Each static point is an independent L-BFGS fit. Unless stated otherwise, each constrained result is based on one mini-batch Adam realization followed by deterministic refinement; the LiCl initialization study instead compares three prescribed warm starts.

\subsection{Physical systems and solver-matched reference paths}

To isolate energy--force selection from representation design, we fix the ACE basis, data partition, component-loss scaling, regularization coefficient, and held-out evaluation within each system before varying the balance. The static route trains an independent L-BFGS model at every $\alpha$ in Eq.~\eqref{eq:alpha_loss}, whereas the constrained route follows Algorithm~\ref{alg:pd_opt} and then refines the model by L-BFGS with $\lambda_{\mathrm{out}}$ fixed. Both routes use Eq.~\eqref{eq:component-loss}, the same ACE basis, and the same held-out split; their fixed-$\gamma$ objectives differ through the regularizer scaling in Eq.~\eqref{eq:lambda_normalized_objective}. We ask whether the constrained route reaches the same held-out error region as its solver-matched static reference.

The liquids test coupled structural and transport responses: molten LiCl probes ionic coordination and Li diffusion, whereas H$_2$O probes O--O structure and molecular diffusion. Si provides a distinct solid-state test, combining a crystalline-to-disordered distribution shift in force prediction with equilibrium, elastic, and vacancy properties. We compare the selected states first in held-out loss coordinates and then through these downstream observables.

The molten-LiCl data of Sivaraman \textit{et al.} \cite{sivaraman2021automated} are divided into 586 training and 52 held-out configurations. The revPBE0-D3 liquid-water data of Cheng \textit{et al.} \cite{cheng2019ab} use a fixed 4:1 partition, giving 1274 training and 319 held-out configurations. The Si split from the dataset of Bart\'ok \textit{et al.} \cite{Bartok2018-fk} instead evaluates a crystalline-to-disordered distribution shift: 2475 crystalline configurations form the training set, whereas the 21 held-out configurations are amorphous and stacking-fault structures. Both held-out structural classes lie outside the crystalline fitting split.

We fit all models with \texttt{ACEpotentials.jl} \cite{witt2023acepotentials}. Appendix~\ref{sec:appendix_protocol_scope} lists the system-specific bases, optimizer settings, and timing conventions that define these comparisons.

\subsection{Protocol dependence of the static scalarization path}
\label{sec:geometry_results}
The static scalarization paths test whether the nominal weight $\alpha$ provides a stable coordinate for the fitted models. Under the specified protocols, the held-out paths develop dominated tails and fail to trace a uniformly improving coordinate. Their knees vary across systems, and their nondominated subsets also change with the solver. We construct the paths in scaled fitting coordinates but report the selected states as physical-unit RMSEs. These finite, protocol-specific sets serve as solver-matched references for the constrained models.

\begin{figure*}[t]
    \centering
    \includegraphics[width=\linewidth]{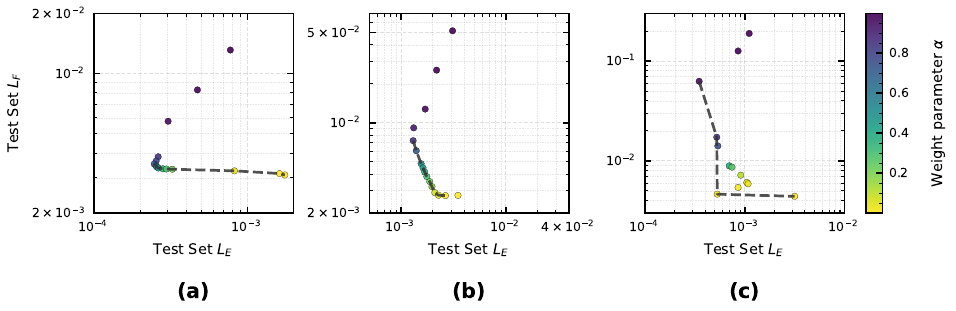}
    \caption{Empirical scalarization paths and nondominated held-out subsets for (a) molten LiCl, (b) liquid H$_2$O, and (c) Si. Each point is an independent L-BFGS fit at fixed $\alpha$, and the grey dashed line connects the nondominated loss pairs. The color scale runs from force-biased ($\alpha\to0$, yellow) to energy-biased ($\alpha\to1$, violet) fitting. Extreme weights produce backward-bending dominated tails, most visibly for LiCl and Si. Both axes are logarithmic. ``Test Set'' denotes the held-out model-selection split defined in Sec.~\ref{sec:formulation}.}
    \label{fig:multidataset_pareto}
\end{figure*}

Figure~\ref{fig:multidataset_pareto} shows the held-out energy--force loss pairs returned by independent L-BFGS fits at the sampled fixed $\alpha$ values. The colored points form the empirical scalarization path, and the grey dashed line connects its nondominated subset. Table~\ref{tab:pareto_rmse} reports physical-unit RMSEs for the heuristic \texttt{ACEpotentials.jl} baseline and the discrete knee selected by Eq.~\eqref{eq:knee_selection}, thereby relating the loss-space geometry to the reported prediction errors. Appendix~\ref{sec:appendix_quantitative} lists additional static reference states, while Appendix~\ref{sec:appendix_knee_sensitivity} gives an alternative knee definition in linear RMSE coordinates.

\begin{table}[htbp]
\centering
\caption{Held-out RMSE for the static \texttt{ACEpotentials.jl} heuristic baseline and the reference knee selected by Eq.~\eqref{eq:knee_selection}. The Selection column gives $\alpha$ for the heuristic rows; the knee rows are selected from the nondominated sampled path. Additional static states are given in Table~\ref{tab:comprehensive_grid_search}.}
\label{tab:pareto_rmse}
\begin{tabular}{l l c c c}
\toprule
\multirow{2}{*}{System} & \multirow{2}{*}{State} & \multirow{2}{*}{Selection} & \multicolumn{2}{c}{Held-out RMSE} \\
\cmidrule(lr){4-5}
 & & & Energy & Force \\
 & & & (meV/atom) & (meV/\AA) \\
\midrule
\multirow{2}{*}{LiCl} & Heuristic & 0.821 & 1.94 & 60.99 \\
                      & Knee & selected & 1.97 & 58.01 \\
\midrule
\multirow{2}{*}{H$_2$O} & Heuristic & 0.610 & 2.79 & 73.26 \\
                        & Knee & selected & 3.31 & 53.67 \\
\midrule
\multirow{2}{*}{Si}   & Heuristic & 0.746 & 4.01 & 118.89 \\
                      & Knee & selected & 4.97 & 68.14 \\
\bottomrule
\end{tabular}
\end{table}

At extreme scalarization weights, the held-out paths bend backward: another sampled model attains lower energy and force errors simultaneously. Increasing the nominal emphasis on one loss can therefore worsen even that held-out coordinate under the stated fitting protocol. Such reversals are consistent with the combined influence of poor conditioning and finite stopping when one data block is weakly weighted.

The nondominated trade-off varies strongly across the tested system--split pairs. For LiCl, the heuristic state already lies near the reference knee: moving to the knee lowers the force RMSE by only $4.9\%$, while the energy RMSE changes by $0.03$ meV/atom. H$_2$O shows a sharper exchange, with a $26.7\%$ reduction in force RMSE accompanied by an $18.6\%$ increase in energy RMSE. The selected knee therefore represents a mild rebalance for LiCl but a pronounced energy--force compromise for water.

The Si comparison is a distribution-shift stress test: its held-out amorphous and stacking-fault structures are absent from the crystalline training split. On these structures, the reference-knee model has a $42.7\%$ lower force RMSE than the heuristic fit, accompanied by a $23.9\%$ higher energy RMSE. This quantified gain characterizes the prescribed distribution-shift split.

Appendix~\ref{sec:appendix_solver} directly probes numerical-protocol dependence. With the scalarization weights and data splits held fixed, replacing the L-BFGS fit by LSQR together with its damping and stopping convention changes the empirical paths, including their nondominated subsets and extreme-weight segments. The comparison probes these numerical choices jointly and establishes the protocol dependence of the returned path. We use the L-BFGS path as a solver-matched reference, with the representation, split, loss scaling, regularization, solver, and stopping rule as components of that reference protocol.

\subsection{Comparison of constrained models with the static reference}
\label{sec:optimization_results}

With the force-loss scale specified, we next test whether a single parameter--multiplier trajectory reaches the solver-matched low-error region. The static calculations play two distinct roles: their low training-force-loss scale sets the numerical magnitude of the prescribed tolerance, whereas the selected static knee provides the held-out comparison state. We evaluate the final constrained model after fixed-$\lambda$ L-BFGS refinement against that knee using held-out energy and force RMSEs in physical units.

All three constrained runs use $\varepsilon_F=10^{-5}$, chosen to match the order of the training-force losses in the low-error static calculations. Because $L_F$ includes configuration weights and package scaling, this value is a protocol-level tolerance in the weighted squared-loss units of Eq.~\eqref{eq:component-loss}. Full-training force-constraint residuals evaluated every $K_{\mathrm{dual}}$ primal steps update $\lambda$ during Adam; the returned $\lambda_{\mathrm{out}}$ is then frozen for L-BFGS refinement of $L_E+\lambda_{\mathrm{out}}L_F+\mathcal{R}$. The reported model is selected by this two-stage procedure and evaluated through its held-out energy and force losses. The associated $\alpha_{\mathrm{eff}}=1/(1+\lambda_{\mathrm{out}})$ specifies the normalized data-term energy fraction. Equation~\eqref{eq:lambda_normalized_objective} shows that regularizer rescaling distinguishes it from the full static-objective coordinate.

\begin{figure*}[t]
    \centering
    \includegraphics[width=\linewidth]{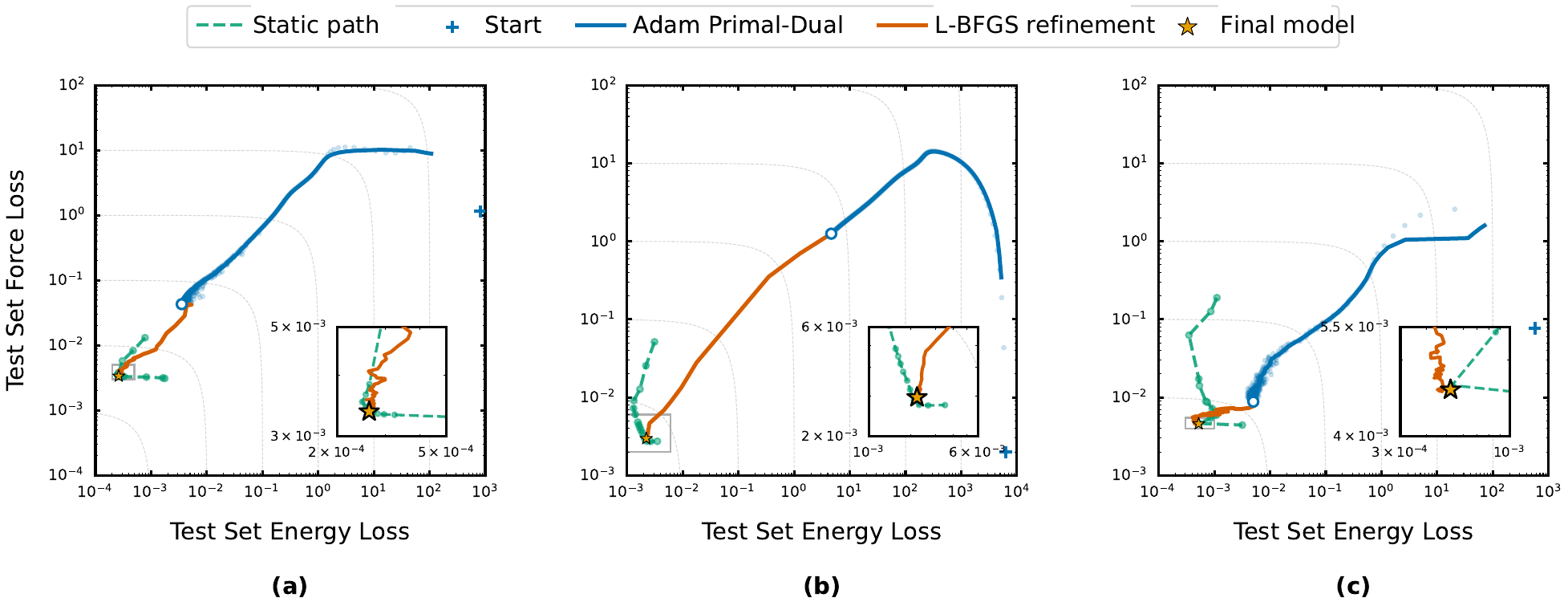}
    \caption{Two-stage constrained trajectories for (a) LiCl, (b) H$_2$O, and (c) Si in held-out energy--force loss space. Each trajectory starts at the cross, reaches the blue open circle after the Adam multiplier iteration, and follows the solid orange segment during fixed-$\lambda$ L-BFGS refinement. The star marks the final returned model. Dashed green curves show the empirical scalarization paths, and the insets resolve the final low-error regions. Both axes are logarithmic.}
    \label{fig:primal_dual}
\end{figure*}

Figure~\ref{fig:primal_dual} projects the two-stage training trajectories into held-out energy--force loss space for comparison with the solver-matched static paths. The Adam segments show coupled parameter--multiplier adaptation through primal steps and periodic residual-driven dual updates, whereas the fixed-$\lambda_{\mathrm{out}}$ L-BFGS segments show deterministic refinement of the full regularized objectives. Training uses the corresponding training objectives, while these held-out coordinates display the optimization histories. Table~\ref{tab:efficiency_metrics} compares the returned models through physical-unit held-out RMSEs and aggregate measured training times.

Across the three systems, $\lambda_{\mathrm{out}}$ spans more than two orders of magnitude, from $1.244$ to $285.62$, while the corresponding normalized energy fraction $\alpha_{\mathrm{eff}}$ decreases from $0.446$ to $0.00349$. These quantities are coordinates of the scaled, system-specific objectives, so their numerical spread remains system and protocol specific.

Table~\ref{tab:efficiency_metrics} reports the measured training cost and final held-out RMSE after L-BFGS refinement. To summarize agreement with the solver-matched static knees, define
\[
    \Delta_{\mathrm{rel}}
    =
    \max_{s\in\{\mathrm{LiCl},\mathrm{H_2O},\mathrm{Si}\}}
    \max_{q\in\{E,F\}}
    \frac{|R_q^{\mathrm{con}}(s)-R_q^{\mathrm{grid}}(s)|}{R_q^{\mathrm{grid}}(s)},
\]
where $R_q^{\mathrm{grid}}$ and $R_q^{\mathrm{con}}$ denote the selected static-knee and final constrained RMSEs, respectively. Across all three systems and both error measures, the constrained RMSEs differ from the static-knee values by at most $2.42\%$. At this accuracy level, the completed sampled grids require $6.7$--$8.9$ times as much measured training time as the constrained fits. These ratios compare each completed grid, including knee refinement, with one constrained run and its refinement; preliminary parameter selection is excluded.

\begin{table}[!t]
\centering
\caption{Aggregate measured training time and held-out RMSE for the sampled static grid and two-stage constrained calculation. Grid RMSEs are those of the solver-matched knee; speedup is calculated from the underlying unrounded times. Energy and force RMSEs are in meV/atom and meV/\AA, respectively. The reported times exclude the preliminary choice of tolerances and optimization parameters.}
\label{tab:efficiency_metrics}
\scriptsize
\setlength{\tabcolsep}{2.2pt}
\begin{tabular}{@{}llcccc@{}} 
\toprule
\multirow{2}{*}{System} & \multirow{2}{*}{Method} & \multirow{2}{*}{Time (s)} & \multirow{2}{*}{Speedup} & \multicolumn{2}{c}{Held-out RMSE} \\
\cmidrule(lr){5-6} 
 & & & & $E$ & $F$ \\
\midrule

\multirow{2}{*}{LiCl} 
 & Static grid & $17\,470$ & -- & 1.97 & 58.01 \\
 & Constrained & $2\,594$  & $6.7\times$ & 1.96 & 58.05 \\
\midrule

\multirow{2}{*}{H$_2$O} 
 & Static grid & $244\,609$ & -- & 3.31 & 53.67 \\
 & Constrained & $27\,521$  & $8.9\times$ & 3.39 & 54.44 \\
\midrule

\multirow{2}{*}{Si} 
 & Static grid & $195\,659$ & -- & 4.97 & 68.14 \\
 & Constrained & $23\,080$   & $8.5\times$ & 4.90 & 66.84 \\
\bottomrule
\end{tabular}
\end{table}

With $\varepsilon_F$ prescribed, the constrained fit adapts $\lambda$ from the force-constraint residual along one parameter--multiplier trajectory. In all three systems, the returned models lie in the solver-matched low-error region. Within the present validation protocol, the static calculations supply the tolerance scale and comparison states, while the two-stage constrained procedure selects the reported models.

\subsection{Dependence on warm-start initialization}

The three-system comparison above initializes each constrained fit from zero coefficients. Warm starts in active learning or dataset expansion instead inherit the energy--force compromise encoded by the preceding fit \cite{vandermause2020fly,jacobs2025practical}.

\begin{figure}[H]
    \centering
    \includegraphics[width=0.90\linewidth]{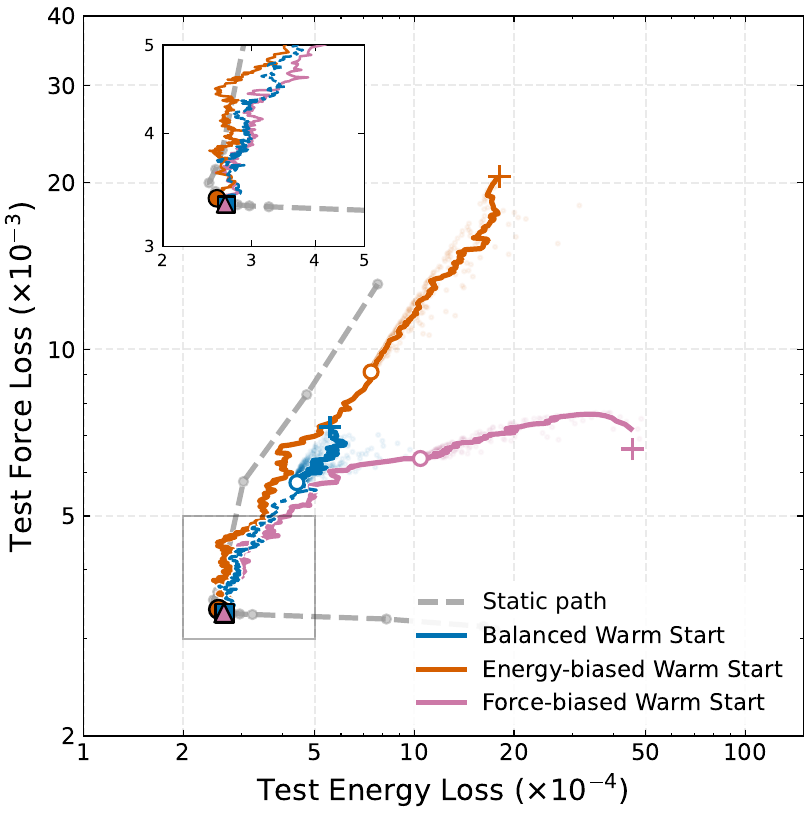}
    \caption{LiCl warm-start trajectories projected into held-out loss space. Energy-biased (orange), force-biased (purple), and intermediate (blue) initial states begin at the crosses. Open circles mark the transition from the Adam multiplier iteration to fixed-$\lambda$ L-BFGS refinement. All three endpoints enter the magnified low-error region of the empirical scalarization path (grey dashed line).}
    \label{fig:licl_warmstart}
\end{figure}

We repeat the LiCl constrained fit from deliberately energy-biased, force-biased, and intermediate warm starts to test how multiplier feedback changes sensitivity to the inherited energy--force compromise. Figure~\ref{fig:licl_warmstart} compares the resulting trajectories and endpoints in held-out loss space.

The three trajectories begin from widely separated energy--force balances but enter a common low-error region of the empirical path. Their distinct endpoints within that region quantify the residual initialization dependence of these LiCl fits.

\subsection{Diamond-Si equilibrium and defect properties}
\label{sec:si_elastic_validation}

We next examine how energy--force selection affects zero-temperature diamond Si. The lattice constant $a_0$ locates the equilibrium geometry, while the elastic constants probe the local curvature of the energy surface. The vacancy formation energy $E_{\mathrm{vac}}$ extends the comparison from near-equilibrium response to point-defect energetics. Together, these observables test whether the reduced force error on the shifted held-out split comes at the expense of equilibrium geometry, elastic response, or defect energetics.

Table~\ref{tab:si_elastic_validation} places the fitted models alongside literature GGA--PBE density functional theory (DFT) values for diamond Si \cite{zongo2024unified}. These independent results set the physical scale for the listed observables. We assess energy--force selection through the relative ordering of the three ACE models under a common property-calculation protocol.

\begin{table}[H]
\centering
\caption{Diamond-Si equilibrium, elastic, and vacancy properties. Literature DFT values are from Ref.~\cite{zongo2024unified}; the other rows are calculations with the heuristic, static-knee, and primal--dual (PD) ACE models.}
\label{tab:si_elastic_validation}
\scriptsize
\setlength{\tabcolsep}{2.2pt}
\begin{tabular}{@{}lcccccc@{}}
\toprule
Model & $a_0$ & $B$ & $C_{11}$ & $C_{12}$ & $C_{44}$ & $E_{\mathrm{vac}}$ \\
 & (\AA) & (GPa) & (GPa) & (GPa) & (GPa) & (eV) \\
\midrule
DFT ref. & 5.4690 & 89.0 & 153.1 & 57.0 & 74.3 & 3.7 \\
Heuristic & 5.47 & 94 & 160 & 61 & 70 & 3.5 \\
Knee & 5.47 & 90 & 155 & 58 & 73 & 3.6 \\
PD & 5.47 & 90 & 154 & 58 & 74 & 3.6 \\
\bottomrule
\end{tabular}
\end{table}

All three models give $a_0=5.47$~\AA{} at the reported precision, so the selected balance has no resolved effect on the equilibrium geometry. Relative to the heuristic model, the static-knee and PD models move all four elastic constants and $E_{\mathrm{vac}}$ toward the corresponding literature values. The PD model thus combines a $42.7\%$ reduction in held-out force RMSE with unchanged equilibrium geometry and systematic shifts in elastic response and vacancy energetics toward the literature scale.

\subsection{Liquid structure and transport properties}
\label{sec:liquid_observable_validation}

For each liquid, we compare the heuristic, static-knee, and PD potentials under a common molecular-dynamics protocol. The density $\rho$ and first-shell position $r_1$ characterize bulk packing and local geometry, $N_1$ probes local coordination, and the diffusion coefficient measures transport. Because replicate uncertainties are unavailable, we interpret the model-to-model differences as directional changes.

Table~\ref{tab:liquid_observable_validation} uses three literature comparators. The LiCl values come from the same molten-LiCl study as the fitting dataset \cite{sivaraman2021automated}. For water, $r_1$ and $N_1$ are experimental diffraction comparators \cite{skinner2014structure}, whereas $\rho$ and $D_{\mathrm{H_2O}}$ are independent 330~K SCAN results \cite{chen2017abinitio}. The water fitting data use revPBE0-D3 \cite{cheng2019ab}, so the SCAN entries serve as external physical scales at a different electronic-structure level.

\begin{table}[H]
\centering
\caption{Liquid structure and transport properties. ``Lit.'' denotes the literature comparator; PD denotes the primal--dual model. LiCl values refer to the 1100~K melt where applicable. The water structural entries are experimental diffraction comparators, whereas its density and diffusion entries are 330~K SCAN results; the latter differ from the revPBE0-D3 electronic-structure level of the fitting data.}
\label{tab:liquid_observable_validation}
\scriptsize
\setlength{\tabcolsep}{2.0pt}
\begin{tabular}{@{}llcccc@{}}
\toprule
System & Observable & Lit. & Heuristic & Knee & PD \\
\midrule
LiCl & $\rho$ (g cm$^{-3}$) & 1.41 & 1.42 & 1.41 & 1.41 \\
LiCl & Li--Cl $r_1$ (\AA) & 2.29 & 2.31 & 2.30 & 2.30 \\
LiCl & Li--Cl $N_1$ & 4.39 & 4.45 & 4.36 & 4.35 \\
LiCl & $D_{\mathrm{Li}}$ ($10^{-4}$ cm$^2$ s$^{-1}$) & 1.74 & 1.5 & 1.7 & 1.7 \\
H$_2$O & $\rho$ (g cm$^{-3}$) & $1.050\pm0.027$ & 1.02 & 1.00 & 1.00 \\
H$_2$O & O--O $r_1$ (\AA) & 2.80 & 2.78 & 2.80 & 2.80 \\
H$_2$O & O--O $N_1$ & 4.3 & 4.6 & 4.4 & 4.4 \\
H$_2$O & $D_{\mathrm{H_2O}}$ (\AA$^2$ ps$^{-1}$) & $0.190\pm0.025$ & 0.15 & 0.18 & 0.18 \\
\bottomrule
\end{tabular}
\end{table}

Across both liquids, $r_1$ changes by at most $0.02$~\AA, whereas $N_1$ and diffusion respond more strongly to the selected balance. The static-knee and PD LiCl models move both coordination and Li diffusion toward the same-source values; for water, $N_1$ approaches the diffraction comparator and $D_{\mathrm{H_2O}}$ enters the listed SCAN range. Water density instead moves farther below the corresponding SCAN range, exposing an observable- and system-specific trade-off.

\section{Discussion and Conclusion}
\label{sec:conclusion}

The central result is that the scientifically relevant object is the fitted model--protocol pair: a nominal energy--force weight identifies a potential only relative to the protocol used for fitting. Dominated and nonmonotone path segments, together with the LSQR comparison, expose this dependence even for a fixed representation.

The force-constrained formulation supplies an explicit selection rule within the specified fitting protocol: minimize the regularized energy loss subject to a prescribed force-loss tolerance. The adaptive stage couples primal descent to projected multiplier updates driven by the constraint residual; the returned multiplier is then frozen to define the regularized objective for deterministic L-BFGS refinement. This two-stage mechanism replaces a scan over nominal weights with one tolerance-controlled parameter--multiplier trajectory followed by a fixed-objective solve.

Across the tested systems, first-shell geometry is least sensitive to balance selection, whereas coordination, transport, and solid-state observables respond more strongly; the water-density trade-off shows that the direction of change remains system specific. This evidence supports force-constrained model selection for the tested fixed-basis ACE problems under the stated protocols. The fitted multiplier and empirical path remain protocol specific, while convergence analysis of the implemented Adam--L-BFGS iteration and tests with learned representations define the next steps toward nonlinear MLIPs.

\section*{Acknowledgment}
During manuscript preparation, the authors used OpenAI Codex for language editing of the abstract and preparation of submission materials. The authors reviewed the resulting text and take responsibility for all submitted content.

\section*{Author Declarations}

\subsection*{Conflict of Interest}
The authors have no conflicts to disclose.

\subsection*{Author Contributions}
Chenyu Wang: Data curation, formal analysis, investigation, software, validation, visualization, writing--original draft. Yangshuai Wang: Conceptualization, formal analysis, methodology, software, validation, writing--original draft, writing--review and editing. Lei Zhang: Conceptualization, methodology, supervision, writing--review and editing.

\section*{Data Availability}
The datasets analyzed in this study are available from the sources cited in the text. Derived data supporting the findings are available from the corresponding author upon reasonable request. The fitting and timing conventions for the stated comparisons are specified in Appendix~\ref{sec:appendix_protocol_scope}.

\appendix

The appendices report the theoretical connection and numerical settings underlying the main-text comparisons. Appendix~\ref{sec:appendix_theory} relates the multiplier update to an exact-gradient result. Appendix~\ref{sec:appendix_protocol_scope} specifies the fitting and timing conventions. Appendices~\ref{sec:appendix_training}--\ref{sec:appendix_solver} report the static references, an alternative knee definition, multiplier time-scale dependence, and the LSQR comparison.

\section{Theoretical Context for the Projected Dual Update}
\label{sec:appendix_theory}

This appendix places Algorithm~\ref{alg:pd_opt} relative to projected gradient descent--ascent theory and states the assumptions required to transfer the exact-gradient result.

\paragraph{Related exact-gradient setting.} Define $\Phi(\boldsymbol{\theta})=\max_{\lambda\in[0,\Lambda_{\max}]}\mathcal{L}_{\mathrm{PD}}(\boldsymbol{\theta},\lambda)$. In the deterministic setting analyzed by Lin \textit{et al.} \cite{lin2020gradient}, the objective is jointly smooth, each primal-variable slice is uniformly Lipschitz, the dual dependence is concave, and the dual domain is convex and bounded. The resulting value function is also assumed to be bounded below. For the present Lagrangian, linearity in $\lambda$ gives concavity, while $[0,\Lambda_{\max}]$ supplies the bounded convex domain. If $\epsilon_{\mathrm{stat}}>0$ is the prescribed stationarity tolerance, exact-gradient descent--ascent with separated primal and dual step sizes has an $\mathcal{O}(\epsilon_{\mathrm{stat}}^{-6})$ deterministic complexity guarantee for Moreau-envelope stationarity of a uniformly selected primal iterate. The guarantee concerns the selected iterate in the capped problem; last-iterate and uncapped KKT behavior lie outside its conclusion.

\paragraph{Connection to the computations.} Applying the exact-gradient result to a quadratic ACE objective requires a bounded coefficient domain. Algorithm~\ref{alg:pd_opt} also evaluates a post-primal-step full-training residual, uses mini-batch Adam and periodic dual updates, applies a prescribed stopping rule, and finishes with fixed-$\lambda$ L-BFGS refinement. The result motivates multiplier projection and time-scale separation, while Sec.~\ref{sec:optimization_results} evaluates the returned last iterates against the static reference values. Convexity of each fixed-$\lambda$ ACE subproblem motivates the deterministic refinement stage.

\section{Fitting Conventions and Timing Scope}
\label{sec:appendix_protocol_scope}

The comparisons in Sec.~\ref{sec:experiments} use fixed data splits, fixed ACE bases within each system, and one timing convention for the static and constrained calculations. The static calculation consists of independent L-BFGS fits at prescribed scalarization weights, using a nine-point logit grid plus knee refinement, a relative objective-change tolerance of $10^{-6}$, and $\gamma=10^{-8}$. The L-BFGS iteration limits are $25\,000$ for LiCl and $10\,000$ for H$_2$O and Si.

For the compact basis notation below, $\bar{\nu}$ is the maximum correlation order, $D_{\max}$ is the maximum total polynomial degree, and $r_{\mathrm{cut}}$ is the cutoff radius. LiCl uses $(\bar{\nu},D_{\max},r_{\mathrm{cut}})=(3,10,5.5~\text{\AA})$; the constrained iteration uses $(\eta_\theta,\eta_\lambda)=(10^{-4},10^{-2})$ and batches of 117 configurations, returning $\lambda_{\mathrm{out}}=1.244$. H$_2$O uses the fixed optimized liquid-water example basis adopted for both calculations; the constrained iteration uses $(2{\times}10^{-5},5{\times}10^{-2})$ and batches of 254 configurations, returning $\lambda_{\mathrm{out}}=13.55$. Si uses $(4,20,6.0~\text{\AA})$; the constrained iteration uses $(10^{-6},3{\times}10^{-1})$ and batches of 618 configurations, returning $\lambda_{\mathrm{out}}=285.62$. All constrained runs use $\varepsilon_F=10^{-5}$, $\lambda_0=0$, $\Lambda_{\max}=10^4$, Adam updates, and $K_{\mathrm{dual}}=50$.

The L-BFGS refinement uses a Hager--Zhang line search initialized from the final Adam iterate, with $\lambda_{\mathrm{out}}$ held fixed. The timing ratios in Table~\ref{tab:efficiency_metrics} compare the aggregate measured training time for the sampled static grid, including knee refinement, with one constrained run and its L-BFGS refinement. The timings exclude preliminary choices of tolerances, learning rates, batch sizes, and stopping criteria, as well as warm-start runs, property calculations, and figure generation. All calculations used Julia v1.11.1, ACEpotentials.jl v0.9.1, Optim.jl v1.13.3, Optimisers.jl v0.3.4, and 16 CPU threads on an Intel Core i7-9750H processor.

The main-text comparison covers the sampled static-weight grid, the ACE heuristic reference, and the primal--dual hybrid under this convention. Deterministic one-dimensional searches over $\lambda$ and other adaptive or multiobjective weighting schemes provide complementary baselines for future matched implementations with the same data, solver, and timing conventions.

\section{Static-Weight References and Multiplier Time Scale}
\label{sec:appendix_training}

\subsection{Static-Weight Reference States}
\label{sec:appendix_quantitative}
The empirical scalarization paths are generated by independent L-BFGS fits at fixed $\alpha$ and then restricted to nondominated held-out loss pairs for knee selection. The heuristic baselines correspond to the data-term ratios $\alpha$ implied by the default energy and force weighting choices in \texttt{ACEpotentials.jl} \cite{witt2023acepotentials}. The selected knee is a discrete reference on the sampled path. Equation~\eqref{eq:knee_selection} first restricts the sampled path to nondominated held-out points and then chooses the maximum-chord-distance point in normalized logarithmic held-out loss space. Table~\ref{tab:comprehensive_grid_search} lists the held-out reference states used in the main text.

\begin{table}[H]
\centering
\caption{Static-weight held-out reference states. ``Energy ext.'' and ``Force ext.'' denote the most energy-biased and force-biased sampled endpoints listed here.}
\label{tab:comprehensive_grid_search}
\scriptsize
\begin{tabular}{llccc}
\toprule
System & State & Weight/selection & $E_{\mathrm{test}}$ & $F_{\mathrm{test}}$ \\
 & & & (meV/atom) & (meV/\AA) \\
\midrule
LiCl & Energy ext. & 0.999 & 3.38 & 114.55 \\
 & Energy mid & 0.850 & 1.97 & 61.90 \\
 & Heuristic & 0.821 & 1.94 & 60.99 \\
 & \textbf{Knee} & Eq.~\eqref{eq:knee_selection} & 1.97 & 58.01 \\
 & Force ext. & 0.001 & 5.08 & 55.73 \\
\midrule
Si & Energy ext. & 0.999 & 7.84 & 435.07 \\
 & Energy mid & 0.850 & 4.45 & 131.47 \\
 & Heuristic & 0.746 & 4.01 & 118.89 \\
 & \textbf{Knee} & Eq.~\eqref{eq:knee_selection} & 4.97 & 68.14 \\
 & Force ext. & 0.001 & 11.50 & 66.32 \\
\midrule
H$_2$O & Energy ext. & 0.999 & 4.02 & 226.83 \\
 & Energy mid & 0.850 & 2.61 & 85.27 \\
 & Heuristic & 0.610 & 2.79 & 73.26 \\
 & \textbf{Knee} & Eq.~\eqref{eq:knee_selection} & 3.31 & 53.67 \\
 & Force ext. & 0.001 & 4.26 & 52.35 \\
\bottomrule
\end{tabular}
\end{table}

The selected knee rows in Table~\ref{tab:comprehensive_grid_search} define the static references used in the main-text comparison. For LiCl, the heuristic state already lies close to the selected knee. H$_2$O and Si show larger force-error reductions from the heuristic state to the knee. Si displays the sharpest trade-off, and its held-out split contains amorphous and stacking-fault structures absent from the crystalline fitting split.

\subsection{Alternative Linear-Scale Knee Definition}
\label{sec:appendix_knee_sensitivity}

The knee rule in Eq.~\eqref{eq:knee_selection} defines the main-text reference by applying the maximum chord distance method \cite{satopaa2011finding} in normalized logarithmic held-out loss space. For comparison, we record an alternative reference definition in physical RMSE coordinates: the nondominated point closest to the utopian corner after linear min--max normalization.

Let $R_E(\alpha)=\mathrm{RMSE}_{E,\mathrm{test}}(\alpha)$ and $R_F(\alpha)=\mathrm{RMSE}_{F,\mathrm{test}}(\alpha)$ denote the physical errors on the held-out set. Within each system-specific nondominated path, define the min--max-normalized linear coordinates
\begin{equation}
    \widetilde R_q(\alpha)
    =
    \frac{R_q(\alpha)-R_q^{\min}}{R_q^{\max}-R_q^{\min}},
    \qquad q\in\{E,F\},
\end{equation}
where the minimum and maximum values are taken over $\alpha\in\mathcal{A}_{\mathrm{nd}}$ for the corresponding system. The reported nondominated paths have nonzero ranges in both coordinates. The linear-scale utopian knee is then
\begin{equation}
    \alpha_{\text{knee}}^{\mathrm{linear}}
    \in
    \arg \min_{\alpha \in \mathcal{A}_{\mathrm{nd}}}
    \left[\widetilde R_E(\alpha)^2+\widetilde R_F(\alpha)^2\right]^{1/2}.
    \label{eq:knee_linear}
\end{equation}
This alternative makes the coordinate-space choice explicit. The canonical knee states in Table~\ref{tab:comprehensive_grid_search} and the static-grid reference points used throughout the main text are selected by Eq.~\eqref{eq:knee_selection}.

\subsection{Time-Scale Dependence of the Primal--Dual Runs}
\label{sec:appendix_ablation}

Batch size and learning rates determine the time-scale separation of the multiplier update. On LiCl, increasing the force-batch size under linear learning-rate scaling reduced stochastic multiplier fluctuations and shortened the time to reach a representative held-out force-RMSE level of 200~meV/\AA. The measured target times were 4054.5, 802.6, and 554.3~s for force batches of 1140, 4580, and 11400 scalar components, respectively.

\begin{figure}[H]
  \centering
  \includegraphics[width=\columnwidth]{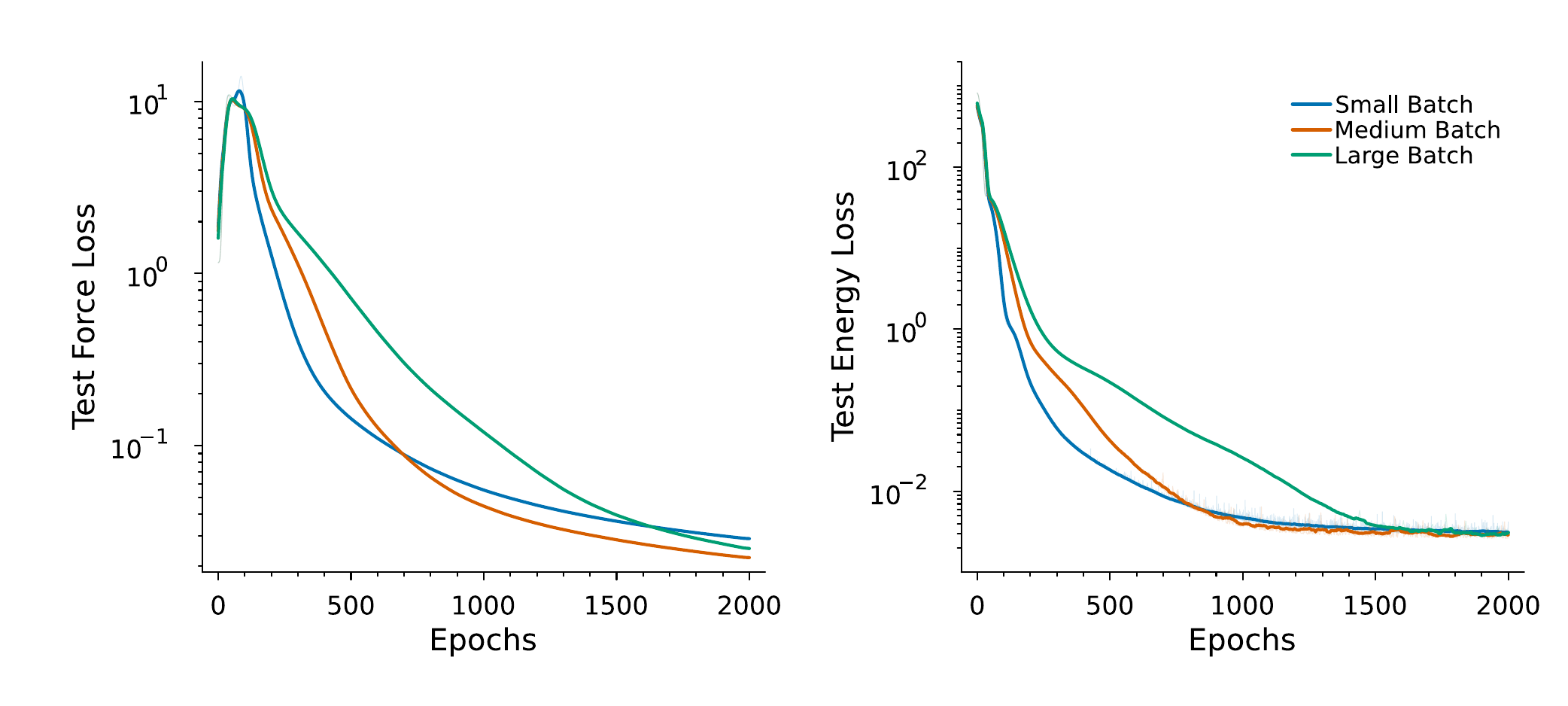}
  \caption{LiCl batch-size dependence during the primal--dual stage. Increasing the force batch from 1140 to 4580 and 11400 scalar components suppresses multiplier-induced fluctuations and shortens the measured time to a held-out force RMSE of 200~meV/\AA. The embedded ``Test'' labels denote the held-out model-selection split.}
  \label{fig:batch_size_ablation}
\end{figure}

\begin{figure*}[t]
    \centering
    \includegraphics[width=\linewidth]{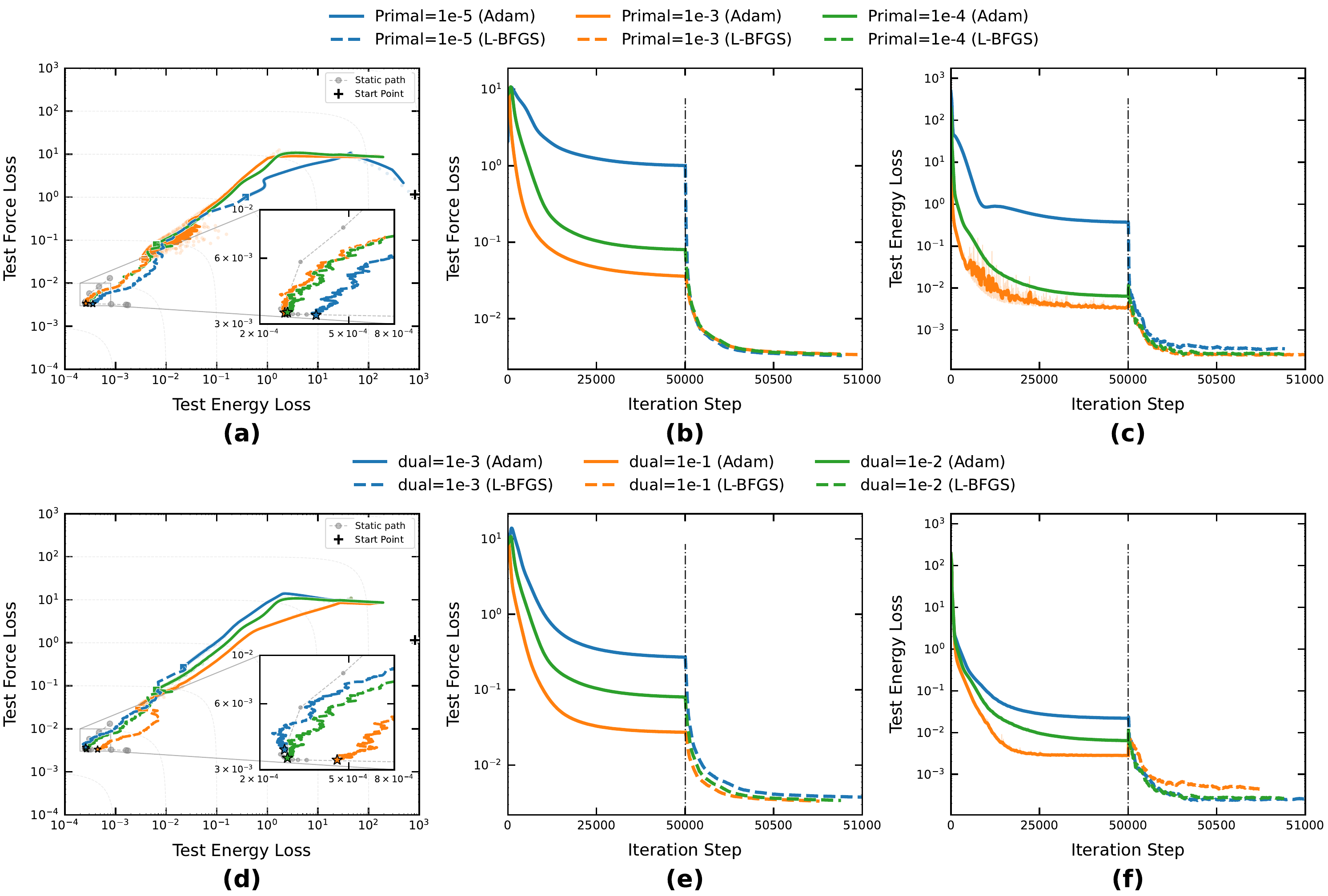}
    \caption{LiCl learning-rate dependence during the primal--dual stage. The panels vary the primal learning rate $\eta_\theta$ and dual learning rate $\eta_\lambda$. Small dual steps delay force correction, overly large dual steps amplify multiplier fluctuations, and the primal step size controls how rapidly the trajectory enters the low-error region. The embedded ``Test'' labels denote the held-out model-selection split.}
    \label{fig:lr_ablation_licl}
\end{figure*}

The LiCl learning-rate sweeps show the same time-scale separation. Small dual steps delay force correction, whereas overly large dual steps produce multiplier fluctuations and can move the endpoint away from the reference knee. Large primal steps move the returned endpoint to different regions of the held-out map, while small primal steps slow the approach to the constrained region. Figure~\ref{fig:lr_ablation_licl} shows the sweep used to identify these regimes.

The useful regime is a two-time-scale compromise: the multiplier must move slowly enough to avoid following mini-batch noise, but quickly enough to correct persistent force residuals before the primal iterate settles. This balance determines the implementation protocol used in the main text.

\section{Dependence of the Empirical Pareto Path on the Linear-Solve Protocol}
\label{sec:appendix_solver}

The LSQR comparison probes the combined linear-solve protocol by changing the solver, damping, and stopping convention while keeping the scalarization weights and data split fixed.

The scalarization weight $\alpha$ is one input to the numerical protocol. The returned model also depends on the linear algebra routine, stopping criterion, regularization, loss normalization, and data split. Thus the map
\[
    \alpha \mapsto
    \bigl(L_{E,\mathrm{test}}(\boldsymbol{\theta}_\alpha),
    L_{F,\mathrm{test}}(\boldsymbol{\theta}_\alpha)\bigr)
\]
depends on the complete fitting protocol, including the chosen solver.

The distinction is most visible near extreme scalarization weights. In those regimes, the weighted fitting problem emphasizes one data block while leaving the other weakly represented in the effective normal equations. Finite-precision arithmetic, iterative stopping criteria, damping, and solver-specific regularization effects can all contribute to which approximate minimizer is returned. L-BFGS and LSQR can therefore trace different held-out paths for the same nominal weighted fitting problem.

\begin{figure*}[!t]
    \includegraphics[width=\linewidth]{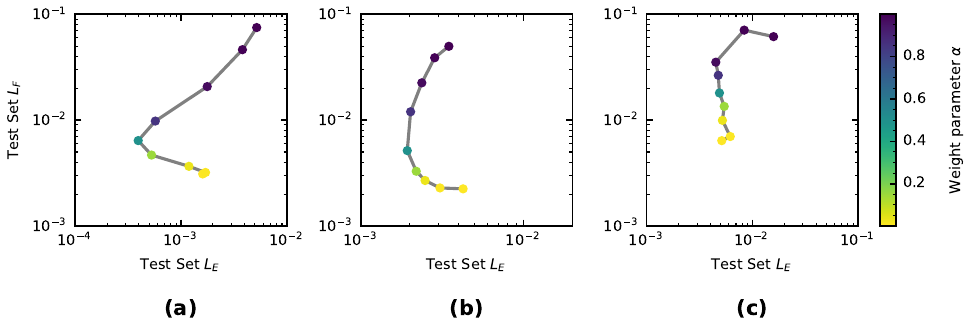}
    \caption{Empirical LSQR scalarization paths for (a) LiCl, (b) H$_2$O, and (c) Si. The paths use the same weights and data splits as the L-BFGS scans but a different solver, damping, and stopping convention. Their nondominated subsets and nonmonotone segments differ from the L-BFGS results, particularly near extreme weights. The main-text L-BFGS path provides the solver-matched reference under the stated protocol.}
    \label{fig:pareto_lsqr}
\end{figure*}

Figure~\ref{fig:pareto_lsqr} shows that the LSQR paths occupy broadly comparable regions of energy--force loss space, but their nondominated subsets and nonmonotone segments differ from the L-BFGS scans. The static grid, knee state, primal--dual endpoint, and aggregate training-time ratios in the main text are therefore specific to the stated L-BFGS protocol.

Consequently, ``$\alpha=0.85$ is balanced'' has meaning only after fixing the data split, loss normalization, regularization, solver, and stopping convention. These choices define the empirical Pareto path and the fitted model.

\bibliographystyle{achemso}
\bibliography{bib}

\end{document}